\documentclass[conference,a4paper]{IEEEtran}

\usepackage{amsmath,amssymb,amsmath,amsfonts}
\usepackage{bm}
\usepackage{bbm}
\usepackage{graphicx}
\usepackage{cite}
\usepackage{epstopdf}
\usepackage{gensymb}
\usepackage{color}
\usepackage{lipsum}
\usepackage{float}
\usepackage{epstopdf}
\usepackage[mathcal]{eucal}
\usepackage{subfiles}
\usepackage[T1]{fontenc}
\usepackage{siunitx}
\usepackage[hyphens]{url}
\usepackage[breaklinks=true]{hyperref}
\usepackage{epsfig,graphics,graphicx,subfig,amssymb,amstext,amsmath,algorithm,algorithmic,wrapfig,multirow}
\usepackage{array}
\usepackage{adjustbox}
\usepackage[dvipsnames]{xcolor}
\usepackage{cite}
\usepackage{times}
\usepackage{placeins}
\usepackage{textcomp}
\usepackage[font=small]{caption}
\usepackage[breaklinks=true]{hyperref}
\usepackage{flushend}
\usepackage{color, colortbl}
\usepackage{tabularx}
\usepackage{bm}
\usepackage{enumerate}
\usepackage{tikz}
\usepackage{pgfplots}
\usepackage{epstopdf}
\usetikzlibrary{shapes,arrows}
\usepackage[utf8]{inputenc}
\usepackage{mathtools}
\usepackage{xcolor}
\usetikzlibrary{chains,arrows,calc,positioning}
\usepackage{amssymb}
\usepackage{pifont}
\usepackage{soul}
\usepackage{breqn} 
\usepackage{booktabs} 
\usepackage{multirow}
\usepackage{enumitem}
\usepackage{tabulary}
\usepackage[normalem]{ulem}
\usepackage{dblfloatfix}
\usepackage{makecell}
\usepackage{lipsum}
\usepackage{amsthm}
\usepackage{comment}

\newtheoremstyle{mystyle}
  {}
  {}
  {\itshape}
  {}
  {\bfseries}
  {.}
  { }
  {}
\theoremstyle{mystyle}

\def\FRF    {{\mathsf{FRF}}}
\def\ISDt   {{\mathsf{ISD}_{\sf T}}}
\def\ISDa   {{\mathsf{ISD}_{\sf A}}}
\def\MNOt   {{\mathsf{MNO}_{\sf T}}}
\def\MNOa   {{\mathsf{MNO}_{\sf A}}}
\def\MNOs   {{\mathsf{MNO}_{\sf S}}}

\newlength \figwidth
\pgfplotsset{compat=1.16}
\definecolor{bittersweet}{rgb}{1.0, 0.44, 0.37}
\definecolor{glaucous}{rgb}{0.38, 0.51, 0.71}
\definecolor{gainsboro}{rgb}{0.86, 0.86, 0.86}
\definecolor{babyblueeyes}{rgb}{0.63, 0.79, 0.95}
\definecolor{silver}{rgb}{0.75, 0.75, 0.75}
\definecolor{neoncarrot}{rgb}{1.0, 0.64, 0.26}
\definecolor{Gray}{gray}{0.9}
\definecolor{LightCyan}{rgb}{0.88,1,1}
\definecolor{BackgroundLightBlue}{rgb}{0.97,0.97,1}
\definecolor{BackgroundGray}{gray}{0.98}

\makeatletter

 \let\oldforeign@language\foreign@language
 \DeclareRobustCommand{\foreign@language}[1]{%
   \lowercase{\oldforeign@language{#1}}}

\def\nb0{{\mathbf{0}}}
\def\nb1{{\mathbf{1}}}

\makeatother

\IEEEoverridecommandlockouts

\begin{document}
\title{Integrating Terrestrial and Non-terrestrial Networks:\\3D Opportunities and Challenges}
\author{
\IEEEauthorblockN{Giovanni Geraci, David L\'{o}pez-P\'{e}rez, Mohamed Benzaghta, and Symeon Chatzinotas}
}

\maketitle

\begin{abstract}
Integrating terrestrial and non-terrestrial networks has the potential of connecting the unconnected and enhancing the user experience for the already-connected,
with technological and societal implications of the greatest long-term significance. 
A convergence of ground, air, and space wireless communications also represents a formidable endeavor for the mobile and satellite communications industries alike, 
as it entails defining and intelligently orchestrating a new 3D wireless network architecture. 
In this article, we present the key opportunities and challenges arising from this (r)evolution by presenting some of its disruptive use-cases and key building blocks, reviewing the relevant standardization activities, and pointing to open research problems. 
By considering two 
multi-operator paradigms, we also showcase how terrestrial networks could be efficiently re-engineered to cater for aerial services, or opportunistically complemented by non-terrestrial infrastructure to augment their current capabilities.
\end{abstract}

\section{Introduction}

A mobile connection is our window to the world. 
The current social, economic, and political drive to reach global wireless coverage and digital inclusion acknowledges connectivity as vital for accessing fair education, medical care, and business opportunities in a post-pandemic society.
Sadly, nearly half of the population on Earth remains unconnected. 
Indeed, rolling out optical fibers and radio transmitters to every location on the planet is not economically viable, 
and reaching the billions who live in rural or less privileged areas has remained a chimera for decades. 
The long-overdue democratization of wireless communications requires a wholly new design paradigm to realize ubiquitous and sustained connectivity in an affordable manner.

Meanwhile, in more urbanized and populated areas, even 5G may eventually fall short of satiating our appetite for mobile internet and new user experiences. 
Life in the 2030s and beyond will look quite different from today’s: 
hordes of network-connected UAVs (uncrewed aerial vehicles) will navigate 3D aerial highways---be it for public safety or to deliver groceries to our doorstep---,
and flying taxis will re-shape how we commute and, in turn, where we live and work. 
The bold ambition of reaching for the sky will take the data transfer capacity, latency, and reliability needs for the underpinning network to an extreme, 
requiring dedicated radio resources and infrastructure for aerial services \cite{geraci2021will,wu20205g}.

In a quest for anything, anytime, anywhere connectivity---
even up in the air---,
next-generation mobile networks may need to break the boundary of the current ground-focused paradigm and fully embrace aerial and spaceborne communications \cite{RinMaaTor2020,giordani2020non}.
To this end, 
the wireless community has already rolled up its sleeves in (re)search for technology enhancements towards a fully integrated terrestrial plus non-terrestrial network (NTN) able to satisfy both ground and aerial requirements.
At first glance, 
terrestrial networks (TNs) could be: 
(i) re-engineered and optimized to support aerial users \cite{MozLinHay2021,chowdhury2021ensuring},
or (ii) complemented by NTN infrastructure such as low Earth orbit (LEO) satellite constellations or aerial base stations (BSs) to further enhance performance \cite{kodheli2020satellite,KarKhoAlf2021}. Cost-related factors may advocate for a progressive roadmap.

In the present paper, we discuss the opportunities and challenges lying behind a 3D integrated TN-NTN.
We begin by providing examples of key use-cases, overviewing the building blocks of an integrated TN-NTN architecture,
and summarizing the most relevant 3GPP standardization activities.
We then introduce the case study of a conventional terrestrial operator pursuing aerial connectivity through two plausible choices: 
(i) deploying dedicated uptilted cells---or partnering with a specialized aerial operator doing so---reusing the same spectrum; 
(ii) leasing infrastructure or solutions from a LEO satellite operator.
We conclude by reviewing the main hurdles that stand in the way to an integrated TN-NTN and pointing out key open problems worthy of further research.

\section{Use-cases, Architecture, and Standardization}

In this section, we describe the main use-cases and components of a plausible integrated TN-NTN, and we summarize the major NTN and UAV standardization advancements.

\subsection{Use-cases}

The opportunities unlocked by integrating TN and NTN capabilities could lead to a vast number of new applications and services. 
In what follows, 
we provide a representative down-selection of the key use-cases. 

\emph{Critical communications:} 
Connectivity from space or air can empower ultra-reliable critical communications in the absence of cellular coverage or during an emergency or natural disaster. In this case, when the ground network becomes dysfunctional and the importance of providing rapid and resilient connectivity cannot be overstated, 
NTNs can ensure replacement coverage through direct access from space/air or even via satellite- or cellular- backhauled UAV radio access nodes.

\emph{Massive IoT and immersive communications:}
NTNs can cover large areas of land or sea populated with both static and nomadic sensory nodes, all collecting real-time data. Aggregating and displaying the latter through AR/VR applications will provide users with spatial and contextual awareness, enabling immersive human-machine interaction, likely one of the 6G killer apps. Depending on the latency requirements and sensory node capabilities, data aggregation could be handled by a LEO constellation in the field of view of a ground gateway or aerial BSs. NTN broadcast/multicast could then pursue content scalability and uninterrupted delivery to users in cars, trains, and vessels.

\emph{Aerial communications:} 
Beyond standalone TNs, 
primarily designed for 2D usage, 
an integrated TN-NTN could support reliable data and control links to multiple UAVs, electrical vertical take-off and landing vehicles (eVTOLs), and aircrafts. 
These services would be guaranteed in specific 3D areas---aerial corridors or waypoint trajectories---where end-devices will be allowed to fly at different heights. The potential of UAVs may only truly be unleashed once the network capabilities and regulations allow for autonomous operation beyond visual line-of-sight (LoS)  \cite{ZenGuvZha2020,SaaBenMoz2020}.

\subsection{Architecture}

\begin{figure}
\centering
\includegraphics[width=\figwidth]{
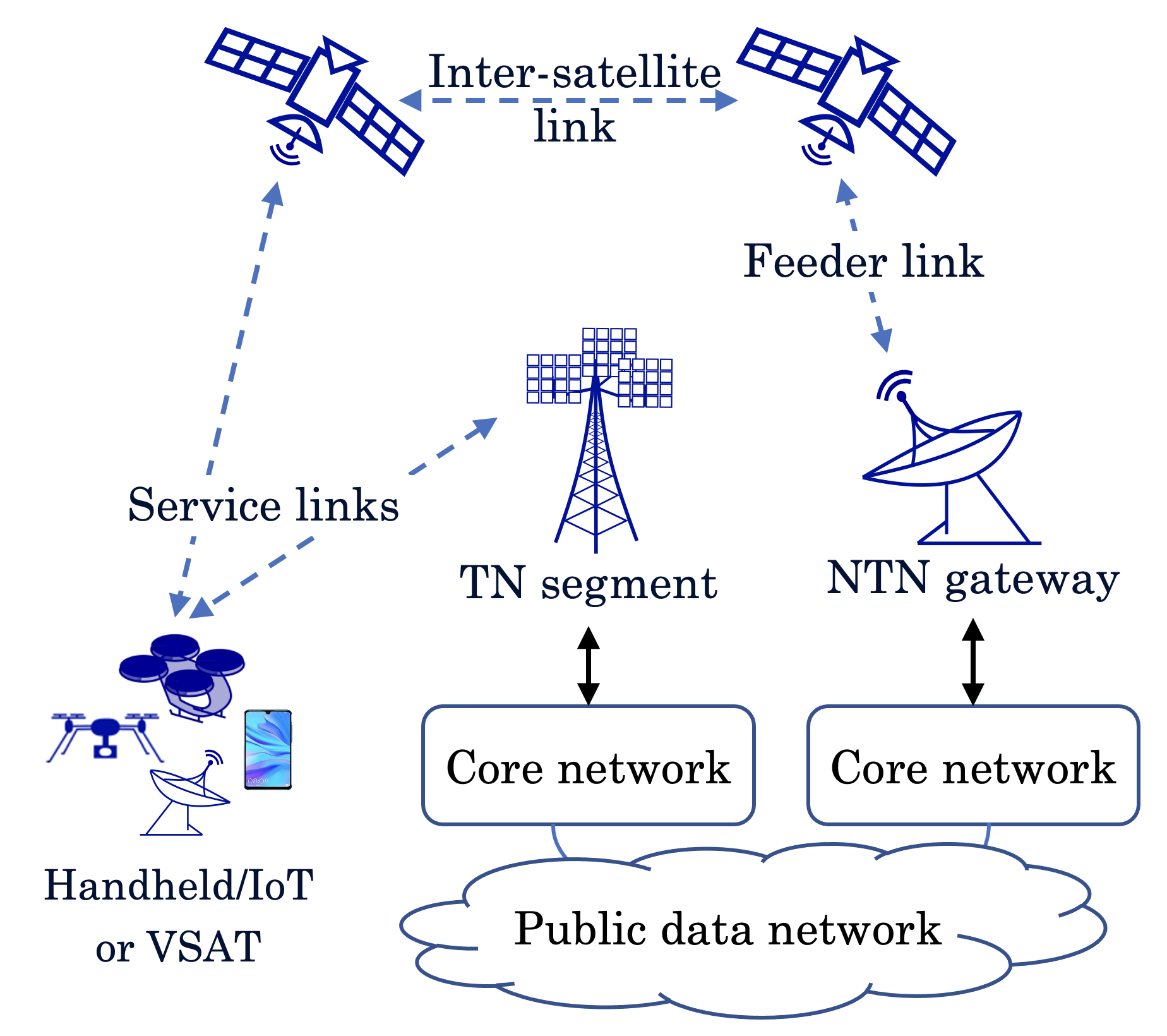}
\caption{Exemplified integrated TN-NTN. 
NTN BS functionalities can be placed onboard satellites or at the NTN gateway, respectively entailing a regenerative or transparent satellite payload \cite{kodheli2020satellite}.}
\label{fig:architecture}
\end{figure}

A simplified integrated TN-NTN architecture is illustrated in Fig.~\ref{fig:architecture}, with service links connecting a user terminal---either handheld/IoT or VSAT---to TN/NTN BSs, feeder links connecting the NTN segment to the ground core network, and (optionally) inter-satellite and/or inter-high-altitude platform stations (HAPS) links.

\subsubsection*{Network platforms} 

The 3D TN-NTN will avail of a multi-layered multi-band infrastructure, arranged hierarchically, with the following nodes operating at different altitudes and offering user-centric coverage and service:
\begin{itemize}
\item 
TN BSs of various size, power, height, and orientation,
operating in sub-6 GHz, mmWave, and eventually THz bands, 
and deployed with different densities. 
Along with conventional downtilted BSs, 
mobile operators may choose to deploy dedicated infrastructure, 
e.g., uptilted cells, to serve aerial users.
\item 
Geostationary orbit (GSO) satellites,
orbiting the equatorial plane at an altitude of about 35786~km, 
and creating fixed beams with a footprint radius of up to 3000 km.
\item 
Non-GSO satellites, 
such as LEO, 
deployed at altitudes between 300–1500~km, 
creating footprints of up to 1000 km radius per beam. 
Unlike their GSO counterpart, 
LEO satellites move fast with respect to a given point on the Earth, 
with an orbital period of just a few hours, 
and thus require large constellations for coverage continuity.
\item 
Aerial BSs such as HAPSs, 
placed in the stratosphere at around 20 km, and
creating multiple cells sized about 10 km each, 
or UAV radio access nodes, 
flying at heights somewhere between 0-1~km.
\item 
Ground gateways connecting aerial and spaceborne platforms to the core network through so-called feeder links.
\end{itemize}

\subsubsection*{Terminals} 

The end-devices of a 3D TN-NTN can be classified as follows:
\begin{itemize}
\item 
Stationary and vehicular ground users (GUEs), 
in areas ranging from dense urban to suburban, rural, and remote.
\item 
UAVs, eVTOLs, and aircrafts, demanding in-flight connectivity at altitudes of few hundred meters, 1--3~km, and 10--12~km, respectively \cite{MozLinHay2021}.
\end{itemize}
Satellite-connected devices can either be handheld/IoT or equipped with a very-small-aperture terminal (VSAT), depending on the use-case and the carrier frequency of the service link. Indeed, the more benign link budget in the S-band (sub-6 GHz) enables direct access to omni- or semi-directional handheld/IoT terminals. 
Operating in the Ka-band (mmWave spectrum) incurs a higher attenuation, which must be compensated with a larger antenna gain by employing a VSAT. The latter can be either fixed or mounted on a moving platform (buses, trains, vessels, or aircrafts), thus giving options for either mobile or fixed broadband access.

\subsection{Standardization}

Standardization work on non-terrestrial communications in 3GPP dates back to 2017 \cite{LinRomEul2021}.
This effort can be classified nowadays into two main areas, namely NTN enhancements and TN support for UAVs. 
The former aims at defining a global standard for future spaceborne communications, fostering an explosive growth in the satellite industry. 
Activities within the latter serve the twofold purpose of ensuring that mobile standards meet the connectivity needs for safe UAV operations, and that other users of the network do not experience a loss of service due to their proximity to UAVs. 
The objectives and outputs of the 3GPP work carried out from Rel-15 up to Rel-17, 
along with the topics currently under study for Rel-18 are outlined as follows and summarized in Table~\ref{tab:3GPP}.
\begin{table*}
\vspace{1mm}
\caption{Overview of the 3GPP standardization work on NTN and UAV communications \cite{LinRomEul2021,geraci2021will,Lin2022}.} 
\label{tab:3GPP}
\centering
\colorbox{BackgroundGray}{%
\begin{tabular}{ |m{0.8cm}|m{7.8cm}|m{7.8cm}| } 
\toprule
\rowcolor{BackgroundLightBlue}
 \textbf{Release} & \textbf{Non-terrestrial Networks (NTN) Enhancements} & \textbf{Support for Uncrewed Aerial Vehicles (UAVs)} \\ \midrule
{Rel-15
}  & 
{\textbf{Study on New Radio (NR) to support NTNs [TR 38.811]}\newline
Identified relevant scenarios for NTN deployment studies and integration in terms of: 
frequency bands (S-band at 2~GHz vs. Ka-band at 10--20~GHz), 
typical footprint sizes and minimum elevation angles, 
antenna models and beam configurations
(Earth-fixed, steered towards a fixed area on the ground, vs. moving beams), 
and NTN terminals (handheld vs. VSAT).
Specified propagation channel models based on TR 38.901 
with NTN-specific modifications.
} &
{\textbf{Enhanced LTE support for aerial vehicles [TR 36.777]}\newline
Proposed user- and network-based solutions for downlink and uplink interference mitigation, mobility, and UAV identification. Specified channel models based on TR 38.901 accounting for the UAV height.

\textbf{Enhancements to measurement report triggering [TS 36.331]}\newline
Enhancements included the addition of two reporting events---H1 (above) and H2 (below) user height thresholds---to help the network identify a UAV and deal with any potential interference. 
}
\\ \midrule
{Rel-16
}  & 
{\textbf{Solutions for NR to support NTNs [TR 38.821]}\newline
Focused on FR1 for handheld and IoT satellite access. 
Identified modifications required for the physical and higher layers and system-level simulation assumptions. Studied the impact of delay on random access, scheduling, and hybrid automatic repeat request (HARQ), as well as mobility management for moving LEO platforms.

\textbf{Using satellite access in 5G [TR 22.822]}\newline
Identified use-cases for the provision of services when considering the integration of 5G satellite-based access components. Identified new services and the corresponding requirements.} &
{\textbf{Remote identification of UAVs [TS 22.825]}\newline
Studied the potential requirements and use-cases for remote UAV identification and the services to be offered accordingly. Aimed at allowing air traffic control and public safety agencies to query the identity and metadata of a UAV and its controller via the UTM to authorize, enforce, and regulate UAV operation.

\textbf{UAV connectivity, identification, and tracking [TR 23.754]}\newline
Considered 3GPP-supported connectivity between UAVs and the UAV traffic management (UTM), as well as the detection and reporting of unauthorized UAVs to the UTM.
}
\\ \midrule
{Rel-17
}  & 
{\textbf{NB-IoT and eMTC support for NTN [TR 36.763]}\newline
Focused on IoT applications by addressing issues related to LTE timing relationships, uplink synchronization, and HARQ.

\textbf{Architecture aspects for using satellite access in 5G [TR 23.737]}\newline
Specified enhancements for RF and physical layer, protocols, radio resource management, and frequency bands. Identified a suitable architecture, addressed TN-NTN roaming and timing-related issues, enhanced conditional handover, and location-based triggering.
}
&
{\textbf{5G enhancements for UAVs [TS 22.125, TS 22.829]}
\newline
Produced new UAV KPIs and communication needs related to: payload, command and control traffic, on-board radio access node (UxNB), and service restrictions and network exposure for the UAV. 

\textbf{Application layer support for UAVs [TR 23.755]}\newline
Studied use-cases for UAV identification and tracking, their impact on the application layer, and UAV-UTM interactions for route authorization, location management, and group communication support. 
}
\\ \midrule
{Rel-18}  & 
{\textbf{NR NTN enhancements}\newline
Will study NR NTN coverage for realistic handheld terminals and access above 10 GHz to fixed and moving platforms. Will investigate requirements for network-verified user location, 
and address mobility and service continuity between TN-NTN and across different NTNs.
} &
{\textbf{NR support for UAVs}\newline 
Will study enhancements on measurement reports, signaling for subscription-based UAV identification and its multicast, additional triggers for conditional handover, and beam management at FR1, including UAV directional antennas and BS uptilt beamforming.
}
\\ \bottomrule 
\end{tabular}
}
\end{table*}

\subsubsection*{NTN enhancements}

In 3GPP parlance, 
the term NTN refers to utilizing satellites or HAPS to offer connectivity services and complement terrestrial networks, especially in remote areas where cellular coverage is unavailable. In Rel-17, 3GPP introduced a set of basic features to enable 5G NR operation over NTNs in FR1, i.e., up to 7.125 GHz. 
3GPP Rel-18 will enhance 5G NR NTN operation by 
improving coverage for handheld terminals, 
studying deployments above 10 GHz, 
addressing mobility and service continuity between TN-NTN as well as across different NTNs,
and investigating regulatory requirements for network-verified user location  \cite{Lin2022}.

\subsubsection*{Support for UAVs}

3GPP introduced 4G LTE support for UAVs back in Rel-15, 
including signaling for subscription-based aerial user identification, 
reporting of UAV height, location, speed, and flight path, 
and new measurement reports to address aerial interference up to a certain density of low-altitude UAVs.
In subsequent releases, 3GPP addressed application layer support and security for connected UAVs, also defining the service interactions between UAVs and the UAV traffic management (UTM) system.
As 5G use-cases evolve, Rel-18 will introduce 5G NR support for devices onboard aerial vehicles,
studying additional triggers for conditional handover, BS uptilting, and signaling to indicate UAV beamforming capabilities, among other enhancements \cite{Lin2022}.



\section{Opportunities}

In this section, we consider two multi-operator case studies to illustrate how terrestrial networks could be 
(i) efficiently re-engineered to support non-terrestrial end-devices such as UAVs, 
or (ii) opportunistically complemented by non-terrestrial infrastructure to augment their current capabilities. 
The main system-level assumptions for these two setups are summarized in Table~\ref{table:parameters}.

\begin{table}
\centering
\caption{System-level parameters. All main assumptions follow 3GPP Technical Reports 38.901, 36.777, 38.811, and 38.821.}
\label{table:parameters}
\def\arraystretch{1.2}
\colorbox{BackgroundGray}{
\begin{tabulary}{\columnwidth}{ |p{2.3cm} | p{5.3cm} | }
\hline
\rowcolor{BackgroundLightBlue}
	\textbf{MNO$_{\mathrm{T}}$ \& MNO$_{\mathrm{A}}$}	&  \\ \hline
  Cell layout				& Hexagonal, 3 sectors/site, 1 BS/sector at $25$~m \\ \hline
  Intersite distance		& $\mathrm{ISD_{T}}\!=\!500$~m, $\mathrm{ISD_{A}}\!=\!\{1500,\!1000,\!500\}$~m \\ \hline
  Frequency band 		& 100~MHz TDD at 3.5~GHz \\ \hline
  Spectrum 			& $\MNOt$ and $\MNOa$ in the same band \\ \hline
  Scheduler & DL: 50~MHz per GUE, 50~MHz per UAV \\ \cline{2-2}
  (round robin) & UL: 10~MHz per GUE, 50~MHz per UAV \\ \hline
  \multirow{2}{*}{Precoding}		& DL: ZF (8 users) or EDA (8 users + 16 nulls)  \\ \cline{2-2} 	 & UL: ZF (4 users) or EDA (4 users + 8 nulls)  \\ \hline
  Downlink power 			& $\MNOt$: 46~dBm, $\MNOa$: 46~dBm or less  \\ \hline   
  Uplink power 			& Fractional power control with $\alpha = 0.80$, $P_{0} = -100$~dBm, and $P_{\textrm{max}}=23$~dBm \\ \hline   
	Antenna elements 		& Horiz./vert. HPBW: $65^{\circ}$, max gain: 8~dBi \\ \hline
	Antenna array 		& $8\times 8$ X-POL, fully digital \\ \hline
	Antenna tilt 		&  $\MNOt$: $12^{\circ}$ (down), $\MNOa$: $-45^{\circ}$ (up) \\ \hline
	Noise figure 			& 7~dB \\ \hline \hline
\rowcolor{BackgroundLightBlue}
	\textbf{MNO$_{\mathrm{S}}$} & \\ \hline
  Cell layout				& Orbit: 600~km, 7 beams centered on a hexagonal grid, elevation angle: variable \\ \hline
  \multirow{2}{*}{Frequency band}		& $\FRF\!\!=\!\!1$: 30+30 MHz (DL+UL) FDD at 2~GHz  \\ \cline{2-2} 	 & $\FRF\!\!=\!\!3$: 10+10 MHz (DL+UL) FDD at 2~GHz  \\ \hline
  Spectrum			& $\MNOs$ and $\MNOt$ in orthogonal bands  \\ \hline
  \multirow{2}{*}{Scheduler}		& DL: single-user round robin, whole band \\ \cline{2-2} 
			& UL: multi-user round robin, $360$~kHz each \\ \hline
  Downlink power			& 34 dBW/MHz per beam  \\ \hline
  Uplink power 			& Always max power $P_{\textrm{max}}=23$~dBm \\ \hline   
  Beam antenna				& Circular aperture, HPBW: 4.41$^{\circ}$, max 30~dBi \\ \hline
  $G/T$				& 1.1 dB/K \\ \hline\hline
\rowcolor{BackgroundLightBlue}
	\textbf{Users} & \\ \hline
  GUE distribution & 15 GUEs per $\MNOt$ cell: 80\% in buildings of 4--8 floors, 20\% outdoor at $1.5$~m   \\ \hline
  UAV distribution 				& 1 UAVs per $\MNOt$ cell, at 150~m \\ \hline
  eVTOL distribution	& 0.1--1 eVTOLs per $\MNOt$ cell, at 1500~m \\ \hline
  Traffic	and load & Full buffer, fully loaded network \\ \hline
  User association				& Based on RSRP (large-scale fading) \\ \hline
  User antenna 		& Omnidirectional, gain: 0~dBi  \\ \hline
  Noise figure 			& 9~dB  \\ \hline 
\end{tabulary}
}
\vspace{-0.5cm}
\end{table}


\subsection{Example I: Re-designing TNs for NTN Terminals}

As the penetration of aerial users increases, a terrestrial mobile network operator (MNO) may choose to cater for UAV connectivity or partner 
with another MNO intending to do so \cite{MozLinHay2021,EAN}. The latter gives rise to the following hypothetical setup with two operators sharing the same spectrum, namely:
\begin{itemize}
    \item 
    A terrestrial operator, $\MNOt$, running a standard network comprised of downtilted cells to serve legacy GUEs.
    \item 
    An aerial operator, $\MNOa$, running a dedicated network of uptilted BSs reserved exclusively for connected UAVs.
\end{itemize}

The deployment sites of both operators are on a hexagonal layout and comprised of three co-located BSs, 
each covering one sector (i.e., a cell) spanning an angular interval of $120^{\circ}$. 
Let $\ISDt$ and $\ISDa$ denote the respective inter-site distances, 
whereby we fix the former to 500~m, and vary the latter to study its effect. 
We assume 15 GUEs for each $\MNOt$ cell, 
and for all values of $\ISDa$, 
we keep the UAV density constant and according to 3GPP Case~3 in TR~36.777, 
yielding \{1,~4,~9\} UAVs/cell under $\ISDa = \{500,~1000,~1500\}$~m, respectively. 
GUEs are located both outdoor at $1.5$~m and indoor in buildings consisting of several floors. 
UAVs fly outdoor at a height of $150$~m. 
We assume all GUEs and UAVs to have a single omnidirectional antenna, 
and to connect to the strongest cell of their respective serving operators. 
Both UAVs and GUEs employ the open-loop power control policy specified in 3GPP TR 36.213. 
The models reported in 3GPP TR 38.901 and TR 36.777 are invoked to characterize the propagation features of all links.

We assume the BSs of $\MNOt$ and $\MNOa$ to be respectively downtilted by $-12^\circ$ and uptilted by $45^\circ$, 
the former being commonplace for $\ISDt = 500$~m, 
and the latter yielding the best UAV performance in most cases. 
Each cell is equipped with an $8 \times 8$ massive MIMO array of cross-polarized semi-directive elements, 
each connected to a separate RF chain, resulting in a total of 128 RF chains. 
For both operators, 
we assume perfect channel state information, and consider two different multi-user precoding paradigms:
\begin{itemize}
\item 
\emph{Zero-forcing (ZF) precoding},
where each BS spatially multiplexes a subset of its users. 
On one hand, 
this paradigm requires low-to-no coordination for radio resource allocation 
since all scheduling, beamforming, and networking decisions are performed individually by each BS. 
On the other hand, such a simplification comes at the cost of inter-MNO co-channel interference.
\item 
\emph{Eigendirection-aware (EDA) precoding},
where BSs dedicate a certain number of spatial degrees of freedom to place radiation nulls, thereby canceling interference on the dominant eigendirections of the inter-cell channel subspace \cite{GarGerLop2019}. 
This approach requires
coordination between $\MNOt$ and $\MNOa$ for channel state information acquisition, 
possibly entailing them to belong to the same network provider.
\end{itemize}

\begin{figure}
\centering
\includegraphics[width=\figwidth]{
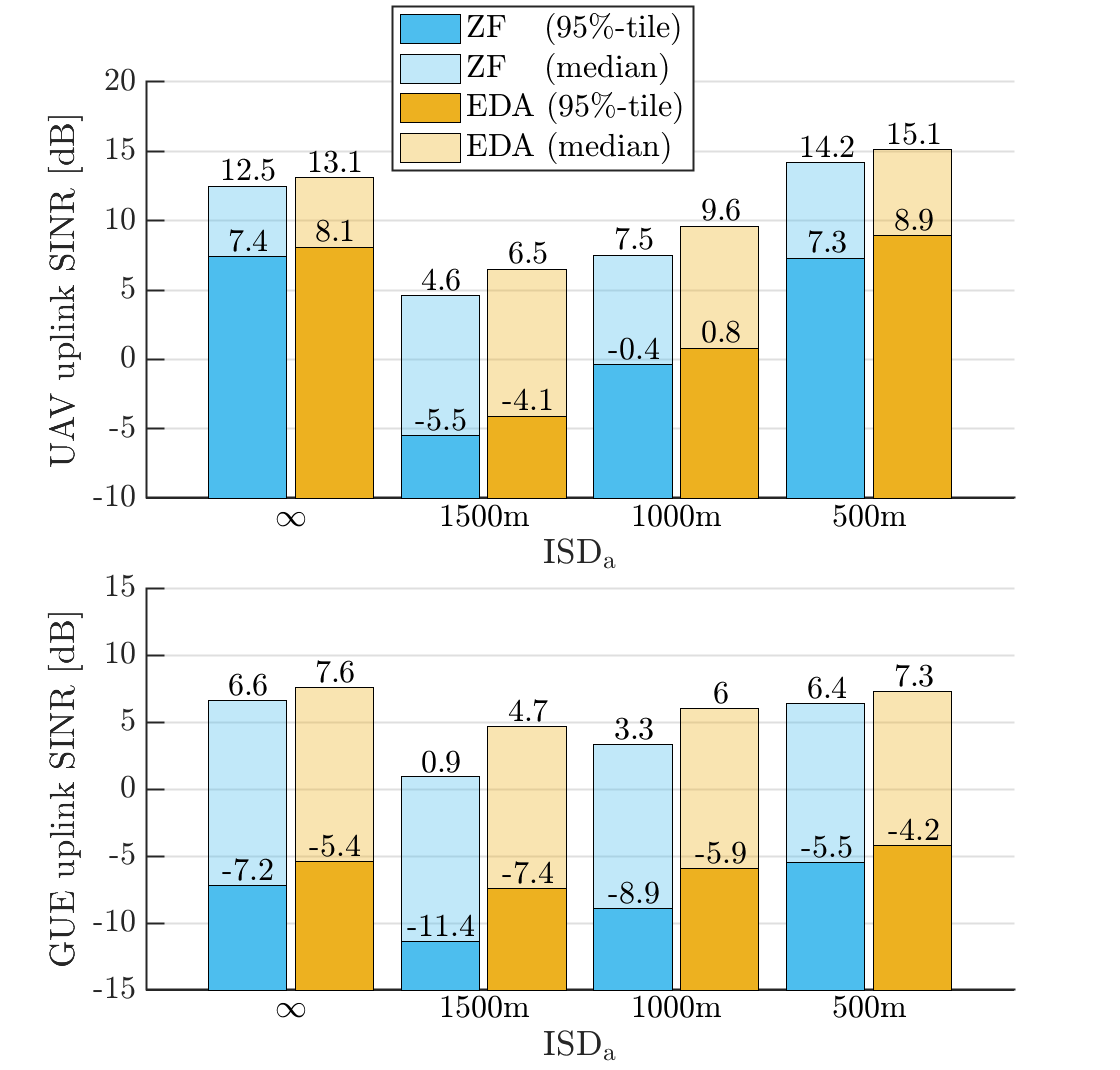}
\caption{Uplink SINR for UAVs (top) and GUEs (bottom) with $\MNOt$ and $\MNOa$ sharing the same spectrum, for $\ISDt=500$~m and a variable $\ISDa$, and employing ZF (blue) or EDA (orange) precoding. $\ISDa=\infty$ denotes all GUEs and UAVs served by standalone $\MNOt$. Solid and transparent bars denote 95\%-tile and median, respectively.}
\label{fig:SINR_MNOA_UL}
\end{figure}

We focus our analysis on the uplink, the more data-hungry direction for UAVs, whose generated transmissions may pose a threat to legacy GUEs \cite{geraci2021will}. Fig.~\ref{fig:SINR_MNOA_UL} shows the SINR attained by UAVs and GUEs for various values of $\ISDa$ and the two precoding schemes. 
These results show the following:
\begin{itemize}
\item 
Offloading UAVs from $\MNOt$ sees their SINR reduced, unless the deployment of $\MNOa$ is sufficiently dense. Importantly, UAVs remain in coverage even under $\ISDa=1500$~m. Offloading, however, provides UAVs with higher data rates, as shown later.
\item
As $\ISDa$ is reduced, 
UAVs are no longer forced to connect to far-off dedicated BSs, and can afford reducing their transmission power and interference generated. This results in an increasing SINR for UAVs and GUEs alike.
\item 
Upgrading from ZF to EDA precoding allows both operators to neutralize the increased intercell interference arising from spectrum sharing. For $\MNOt$, this countermeasure is key to preserve the legacy GUEs performance.
\end{itemize}

While not shown for space constraints, similar observations can be made for the downlink, with two caveats:
\begin{itemize}
    \item 
    UAVs turn from originators to main victims of inter-MNO interference. 
    Reducing $\ISDa$ allows $\MNOa$ a corresponding power reduction, which can be used to trade off UAVs and GUEs performance.
    \item 
    The benefits of EDA nullsteering are mostly confined to---and needed by---$\MNOa$. 
    Indeed, the dominant channel eigendirections for both operators correspond to users most vulnerable to downlink interference. 
    Intuitively, in the presence of UAVs, 
    their strong LoS channels dominate said subspace and most nulls target receiving UAVs.
\end{itemize}

\begin{figure}
\centering
\includegraphics[width=\figwidth]{
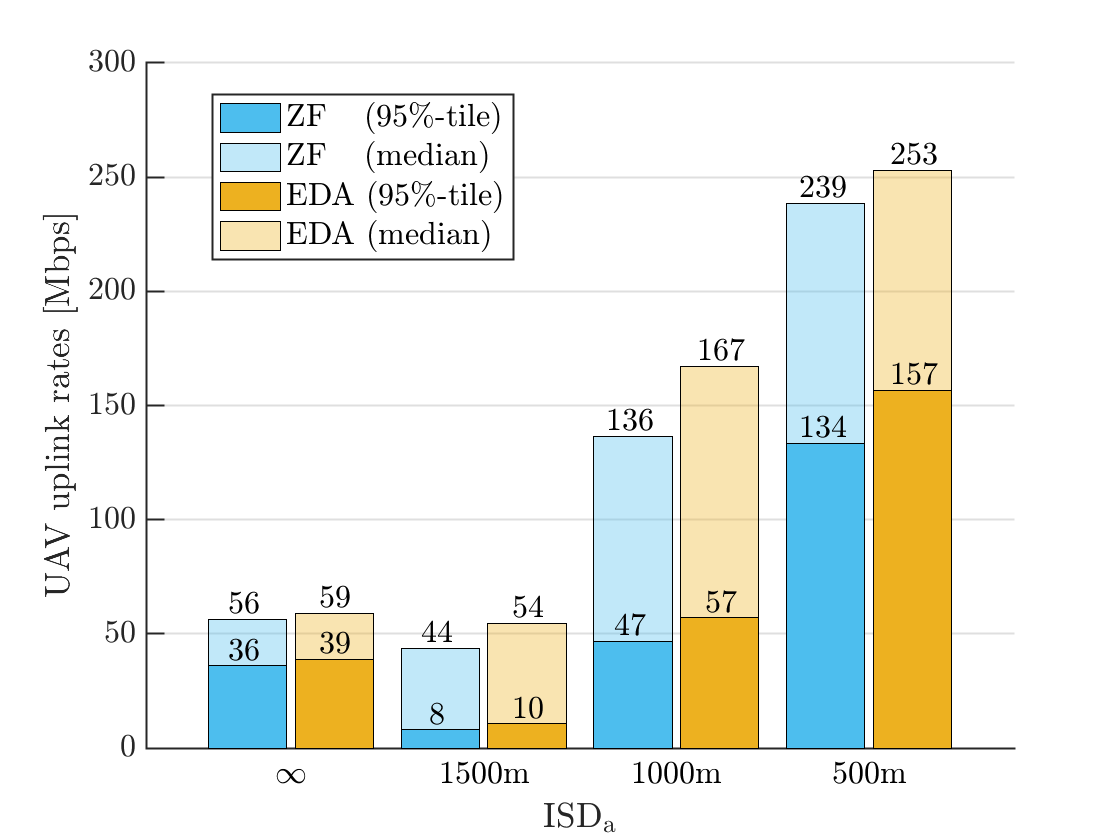}
\caption{Uplink UAV rates with $\MNOt$ and $\MNOa$ sharing the same spectrum, for $\ISDt=500$~m and a variable $\ISDa$, and employing ZF (blue) or EDA (orange) precoding. $\ISDa=\infty$ denotes all GUEs and UAVs served by standalone $\MNOt$. Solid and transparent bars denote 95\%-tile and median, respectively.}
\label{fig:rate_MNOA_UL}
\end{figure}

Under the right deployment and interference mitigation choices, the dual-MNO paradigm can offer comparable SINRs to a setup where GUEs and UAVs are all served by $\MNOt$. However, the spatial and spectrum reuse gains provided by $\MNOa$ reflect in the UAV data rates, reported in Fig.~\ref{fig:rate_MNOA_UL} for the uplink. These largely benefit from increasing the deployment density of $\MNOa$ and employing EDA precoding. Focusing on the 95\%-tile, standalone $\MNOt$ with ZF provides 36~Mbps as opposed to the 134~Mbps achievable with $\MNOt$-plus-$\MNOa$ and $\ISDa=500$~m. 
The former may be sufficient for remote UAV controlling through HD video, whereas the latter may also empower 8K real-time video live broadcast (for future VR applications) and $4\!\times\!4$K AI surveillance (for control and anti-collision in building-intensive areas, lacking positioning accuracy) \cite{geraci2021will}.

\subsection{Example II: Complementing TNs with NTN Infrastructure}

While primarily targeting underserved areas, 
NTNs may also be leveraged to augment urban connectivity, 
e.g., with $\MNOt$ opportunistically leasing spectrum and infrastructure from a satellite service provider. 
In this example, we study the benefits of such an arrangement when offering service to passengers onboard eVTOLs, flying at 1500~m over an urban area \cite{MozLinHay2021}. Let us define:
\begin{itemize}
    \item 
    The same operator $\MNOt$ as in Example I.
    \item 
    A satellite operator $\MNOs$, 
    availing of a LEO constellation 
    and operating in an orthogonal S-band (sub-6~GHz).
\end{itemize}

Each LEO BS of $\MNOs$ generates multiple Earth-moving beams pointing to the ground in a hexagonal fashion, 
each creating one corresponding NTN cell \cite{SedFelLin2020}. 
Due to its orbital movement, 
the LEO satellite may be seen by the users under a variable elevation angle, 
defined as the angle between the line pointing towards the satellite and the local horizontal plane, 
whereby angles closer to $90^{\circ}$ yield shorter LEO-to-user distances, and are more likely to be in LoS.
Besides the elevation angle, the NTN performance is affected by the beam frequency reuse factor (FRF).
With $\FRF = 1$, 
all frequency resources are fully reused across all beams, 
whereas with $\FRF = 3$, 
they are partitioned into three sets, 
each reused every three beams. 
The assumptions reported in 3GPP TR 38.811 and 38.821 are used to characterize the main NTN propagation features.

\begin{figure}
\centering
\includegraphics[width=\figwidth]{
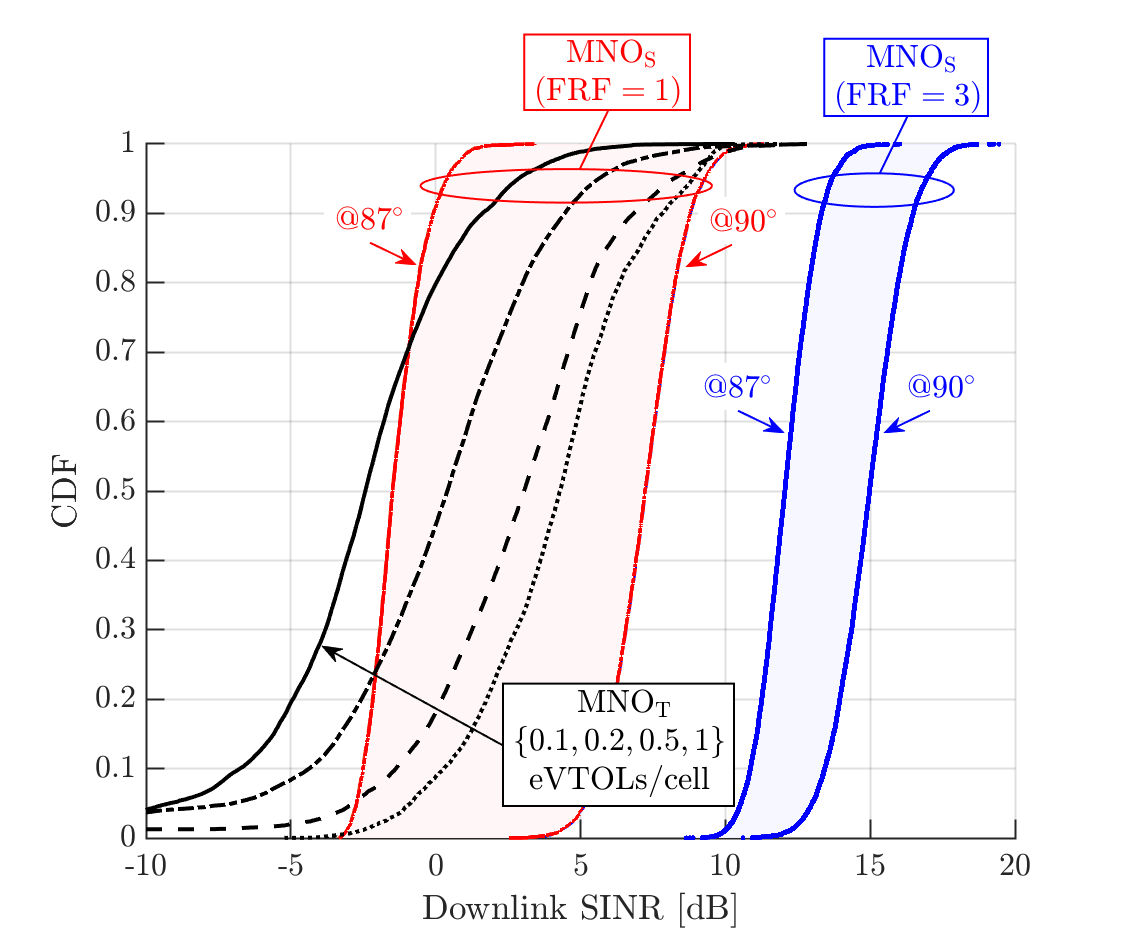}
\caption{Downlink SINR for eVTOL passengers when connected to $\MNOt$ and when offloaded to $\MNOs$. For the latter, various LEO satellite elevation angles and beam FRFs are considered.}
\label{fig:SINR_DL_MNOS}
\end{figure}

This time we focus on the downlink, likely the predominant direction for eVTOL occupants. For the latter, Fig.~\ref{fig:SINR_DL_MNOS} shows the CDF of the downlink SINR experienced when all are served by $\MNOt$ and when their traffic is offloaded to $\MNOs$. For $\MNOs$, various LEO elevation angles are considered. The following remarks can be made:
\begin{itemize}
    \item 
    A standalone $\MNOt$ employing ZF struggles to guarantee coverage to eVTOLs as they proliferate. Indeed, increasing their number from 0.1 to 1 per cell incurs a progressively larger outage, i.e., SINR~$<-5$~dB, reaching up to $18\%$ of the cases (solid black). This is due to the insufficient angular separation between users, caused by their density and sheer height, which also renders nullsteering (not shown) unhelpful.
    \item
    Offloading traffic from $\MNOt$ to $\MNOs$ yields universal coverage with SINRs ranging between $-3$~dB and $17$~dB for the elevation angles and beam FRFs considered.
    \item 
    Moving from $\FRF=3$ to $\FRF = 1$ entails full reuse---and thus inter-beam interference---, degrading the median downlink SINR by approximately $8$~dB and $14$~dB for elevation angles of $90^{\circ}$ and $87^{\circ}$, respectively.
    \item 
    The SINR experiences a prominent degradation when the LEO satellite moves from $90^{\circ}$ to $87^{\circ}$, 
    owing to a larger propagation distance and a lower antenna gain, with the median loss in excess of 8~dB for $\FRF=1$.
    Nonetheless, all offloaded users still remain in coverage, even in the presence of inter-beam interference ($\FRF = 1$).
\end{itemize}

As for the achievable rates, assuming one eVTOL passenger per cell over an area of $\SI{10.8}{km^2}$---the size of Sant Mart\'{i}, Barcelona's business district---yields a total of 150 users, out of which those in outage ($18\%$, i.e., 27 users) could be offloaded to $\MNOs$. Under an ideal elevation angle of $90^{\circ}$ and $\FRF=3$, they would experience median rates of 3~Mbps. Reducing the density of eVTOLs rapidly increases their experienced rates as both their absolute number shrinks and so does the outage percentage from $\MNOt$. Specifically, 0.5 and 0.2 eVTOLs per cell respectively yield 75 and 30 eVTOLs in total. Out of these, 8.8\% and 2.6\% experience SINRs below $-5$~dB, for a total of 7 and 1 eVTOLs incurring outage, respectively. When offloaded to $\MNOs$, their median rates would be of around 11~Mbps and 80~Mbps, respectively.

While our findings are encouraging, they also suggests that in a future with hordes of high-altitude vehicles, broadband aerial communications may require higher NTN spatial reuse through narrow beams and possibly operating in the Ka-band \cite{giordani2020non,SedFelLin2020}. This option may be viable for relayed access through a more directive receiver mounted onboard the eVTOL \cite{EAN}.



\section{Challenges and Research Directions}

The availability of TN plus NTN segments is a prerequisite for realizing a 3D wireless network. 
Jointly and optimally designing and operating all platforms and nodes requires further disruptive and interdisciplinary research. 
In this section, 
we identify the key obstacles that stand in the way along with the most needed technological enablers.

\subsection{The Challenge of Extreme Heterogeneity}

One chief challenge in realizing an integrated TN-NTN arises from its extreme heterogeneity, 
reflected at different levels as outlined below. 

\subsubsection*{Radio propagation features}
NTNs comprise systems and end-devices at different altitude layers, each with own service features. 
For instance, GEO satellites provide stable and continuous links to ground devices with a considerable propagation delay, 
whereas LEO satellites are characterized by lower-delay interfaces, but may suffer from service discontinuity depending on the constellation density. The type of service provided by each layer must be mapped to the user demand, factoring in the interplay of different layers, through dynamic TN-NTN quality of experience management and scheduling.  

\subsubsection*{Node and device capabilities}
By design, GEO satellites differ from LEO satellites in terms of redundancy mechanisms, antenna designs, transceivers, operational frequency, and/or internal resources (e.g., storage, processing, and power availability). The variance in capabilities is yet more apparent with aerial vehicles, as they are conceived for largely different purposes and environments, and terminals, whose antennas range from small and isotropic to active ones capable of tracking. The above further exacerbates the need for network management, 
to guarantee a near-optimal use of radio resources while leveraging this heterogeneity.

\subsubsection*{Ownership and operations} 
Mega-constellations are emerging to expand Internet coverage through hundreds or thousands of satellites, bringing about frequency coordination and collision avoidance issues, among others. 
While current systems lack interoperability, with each operator featuring a vertically integrated stack, 3GPP standardization will be crucial for interconnection, 
giving way to more heterogeneous scenarios. With multiple systems designed and operated in an ad-hoc fashion, their decentralized management and optimization may be a cornerstone to realizing a practical integrated TN-NTN. 


\subsection{Research Directions}

Its extreme heterogeneity makes realizing a 3D network a remarkable endeavor. In the sequel, we propose much-needed research towards an integrated TN-NTN \cite{geraci2021will,kodheli2020satellite}.

\subsubsection*{3D radio access}
Next-generation networks will have to connect flying end-devices at all heights, including their occupants. Our preliminary results vouch for exploiting dedicated uptilted cells and NTN platforms to support aerial services. Nonetheless, operators will have to seek optimal performance-cost tradeoffs, ensuring coexistence between aerial and legacy ground users, and between different co-channel technologies. This goal calls for sophisticated interference management schemes leveraging time, frequency, power, and spatial degrees of freedom, and designed atop realistic air-to-ground channel models. 

\subsubsection*{3D mobility management and multi-connectivity}

Integrated TN-NTN will face the upcoming and unprecedented mobility challenges brought about both by flying end-devices and by a mobile infrastructure, dynamically dealing with user cell selection, re-selection, and configuration. Beyond current power-triggered procedures, 
novel use-case-specific and asymmetric approaches will be required, also accounting for the handover direction, e.g., within a vertical layer (within a LEO constellation or inter-HAPS) or across technologies (ground-to-air/space or vice versa). Optimal mobility management policies will need to trade off reliability, spectral- and energy-efficient load balancing, and signaling overhead caused by conditional handover preparations and radio link failures.

\subsubsection*{3D network management and orchestration}

Meeting the heterogeneous and ever more stringent traffic needs across a 3D wireless network will require optimal load distribution, defining the slices of radio resources to be assigned to each service class, accounting for the features of the available TN/NTN radio links, and following their rapidly varying topology. A service orchestrator should dynamically allocate resources at NTN nodes according to their availability, mobility patterns, architecture hierarchy, and incoming traffic, ensuring seamless service continuity to the end-user in spite of intermittent service link availability and feeder link disruptions. Besides communications, computation and caching resources scattered across TN and NTN nodes will also need to be optimally allocated and leveraged.

\section{Conclusion}

In this paper, we connected the dots between ground, aerial, and spaceborne communications, and reviewed the key opportunities and challenges brought about by integrating terrestrial and non-terrestrial networks. 
We studied augmenting a ground deployment with uptilted cells, and also complementing it with a LEO constellation. We found both to be promising avenues for supporting aerial communications, under the right design choices: the former entails advanced interference mitigation capabilities, the latter hinges on a sufficiently dense constellation---to guarantee near-zenith coverage---and a carefully designed beam reuse.

\typeout{}
\bibliographystyle{IEEEtran}
\bibliography{journalAbbreviations,bibl}

\section*{Biographies}
\small

\noindent
\textbf{Giovanni Geraci} is an Assistant Professor at Univ. Pompeu Fabra in Barcelona. He is an IEEE ComSoc Distinguished Lecturer, co-edited the book ``UAV Communications for 5G and Beyond'', and received the IEEE ComSoc EMEA Outstanding Young Researcher Award.

\vspace{0.2cm}
\noindent
\textbf{David L\'{o}pez-P\'{e}rez} is an Expert and Technical Leader at Huawei Research in Paris. He was a Bell Labs Distinguished Member of Technical Staff and has co-authored 150+ research articles, 50+ filed patents, and two books on small cells and ultra-dense networks.

\vspace{0.2cm}
\noindent
\textbf{Mohamed Benzaghta} is a Ph.D. candidate at Univ. Pompeu Fabra in Barcelona. He received B.Sc. and M.Sc. degrees from Atilim Univ. in Ankara and his research interests include the integration of terrestrial and non-terrestrial wireless communications.

\vspace{0.2cm}
\noindent
\textbf{Symeon Chatzinotas} is Full Professor and Head of the SIGCOM Research Group at SnT, University of Luxembourg, where he is acting as a PI for more than 20 projects.
He was the co-recipient of the 2014 IEEE Distinguished Contributions to Satellite Communications Award and has co-authored more than 450 technical papers.

\end{document}